\documentclass[twocolumn,nofootinbib,showpacs,prl,superscriptaddress,secnumarabic,amssymb,nobibnotes,aps,floatfix]{revtex4}
\usepackage{graphicx}
\usepackage{hyperref}
\usepackage{amsmath}
\usepackage{dcolumn}
\usepackage{bm}

\begin{document}

\title{First Exclusive Measurement of Deeply Virtual Compton Scattering off $^4$He: Toward the 3D Tomography of Nuclei}

\newcommand*{\ANL}{Argonne National Laboratory, Argonne, Illinois 60439}
\newcommand*{\ANLindex}{1}
\affiliation{\ANL}
\newcommand*{\ASU}{Arizona State University, Tempe, Arizona 85287-1504}
\newcommand*{\ASUindex}{2}
\affiliation{\ASU}
\newcommand*{\CSUDH}{California State University, Dominguez Hills, Carson, CA 90747}
\newcommand*{\CSUDHindex}{3}
\affiliation{\CSUDH}
\newcommand*{\CANISIUS}{Canisius College, Buffalo, NY}
\newcommand*{\CANISIUSindex}{4}
\affiliation{\CANISIUS}
\newcommand*{\CMU}{Carnegie Mellon University, Pittsburgh, Pennsylvania 15213}
\newcommand*{\CMUindex}{5}
\affiliation{\CMU}
\newcommand*{\CUA}{Catholic University of America, Washington, D.C. 20064}
\newcommand*{\CUAindex}{6}
\affiliation{\CUA}
\newcommand*{\SACLAY}{Irfu/SPhN, CEA, Universit\'e Paris-Saclay, 91191 Gif-sur-Yvette, France}
\newcommand*{\SACLAYindex}{7}
\affiliation{\SACLAY}
\newcommand*{\CNU}{Christopher Newport University, Newport News, Virginia 23606}
\newcommand*{\CNUindex}{8}
\affiliation{\CNU}
\newcommand*{\UCONN}{University of Connecticut, Storrs, Connecticut 06269}
\newcommand*{\UCONNindex}{9}
\affiliation{\UCONN}
\newcommand*{\FU}{Fairfield University, Fairfield CT 06824}
\newcommand*{\FUindex}{10}
\affiliation{\FU}
\newcommand*{\FERRARAU}{Universita' di Ferrara , 44121 Ferrara, Italy}
\newcommand*{\FERRARAUindex}{11}
\affiliation{\FERRARAU}
\newcommand*{\FIU}{Florida International University, Miami, Florida 33199}
\newcommand*{\FIUindex}{12}
\affiliation{\FIU}
\newcommand*{\FSU}{Florida State University, Tallahassee, Florida 32306}
\newcommand*{\FSUindex}{13}
\affiliation{\FSU}
\newcommand*{\Genova}{Universit$\grave{a}$ di Genova, 16146 Genova, Italy}
\newcommand*{\Genovaindex}{14}
\affiliation{\Genova}
\newcommand*{\GWUI}{The George Washington University, Washington, DC 20052}
\newcommand*{\GWUIindex}{15}
\affiliation{\GWUI}
\newcommand*{\ISU}{Idaho State University, Pocatello, Idaho 83209}
\newcommand*{\ISUindex}{16}
\affiliation{\ISU}
\newcommand*{\INFNFE}{INFN, Sezione di Ferrara, 44100 Ferrara, Italy}
\newcommand*{\INFNFEindex}{17}
\affiliation{\INFNFE}
\newcommand*{\INFNFR}{INFN, Laboratori Nazionali di Frascati, 00044 Frascati, Italy}
\newcommand*{\INFNFRindex}{18}
\affiliation{\INFNFR}
\newcommand*{\INFNGE}{INFN, Sezione di Genova, 16146 Genova, Italy}
\newcommand*{\INFNGEindex}{19}
\affiliation{\INFNGE}
\newcommand*{\INFNRO}{INFN, Sezione di Roma Tor Vergata, 00133 Rome, Italy}
\newcommand*{\INFNROindex}{20}
\affiliation{\INFNRO}
\newcommand*{\INFNTUR}{INFN, Sezione di Torino, 10125 Torino, Italy}
\newcommand*{\INFNTURindex}{21}
\affiliation{\INFNTUR}
\newcommand*{\ORSAY}{Institut de Physique Nucl\'eaire, CNRS/IN2P3 and Universit\'e Paris Sud, Orsay, France}
\newcommand*{\ORSAYindex}{22}
\affiliation{\ORSAY}
\newcommand*{\ITEP}{Institute of Theoretical and Experimental Physics, Moscow, 117259, Russia}
\newcommand*{\ITEPindex}{23}
\affiliation{\ITEP}
\newcommand*{\JMU}{James Madison University, Harrisonburg, Virginia 22807}
\newcommand*{\JMUindex}{24}
\affiliation{\JMU}
\newcommand*{\KNU}{Kyungpook National University, Daegu 41566, Republic of Korea}
\newcommand*{\KNUindex}{25}
\affiliation{\KNU}
\newcommand*{\LPSC}{LPSC, Universit\'e Grenoble-Alpes, CNRS/IN2P3, Grenoble, France}
\newcommand*{\LPSCindex}{26}
\affiliation{\LPSC}
\newcommand*{\MISS}{Mississippi State University, Mississippi State, MS 39762-5167}
\newcommand*{\MISSindex}{27}
\affiliation{\MISS}
\newcommand*{\UNH}{University of New Hampshire, Durham, New Hampshire 03824-3568}
\newcommand*{\UNHindex}{28}
\affiliation{\UNH}
\newcommand*{\NSU}{Norfolk State University, Norfolk, Virginia 23504}
\newcommand*{\NSUindex}{29}
\affiliation{\NSU}
\newcommand*{\OHIOU}{Ohio University, Athens, Ohio  45701}
\newcommand*{\OHIOUindex}{30}
\affiliation{\OHIOU}
\newcommand*{\ODU}{Old Dominion University, Norfolk, Virginia 23529}
\newcommand*{\ODUindex}{31}
\affiliation{\ODU}
\newcommand*{\URICH}{University of Richmond, Richmond, Virginia 23173}
\newcommand*{\URICHindex}{32}
\affiliation{\URICH}
\newcommand*{\ROMAII}{Universita' di Roma Tor Vergata, 00133 Rome Italy}
\newcommand*{\ROMAIIindex}{33}
\affiliation{\ROMAII}
\newcommand*{\MSU}{Skobeltsyn Institute of Nuclear Physics, Lomonosov Moscow State University, 119234 Moscow, Russia}
\newcommand*{\MSUindex}{34}
\affiliation{\MSU}
\newcommand*{\SCAROLINA}{University of South Carolina, Columbia, South Carolina 29208}
\newcommand*{\SCAROLINAindex}{35}
\affiliation{\SCAROLINA}
\newcommand*{\TEMPLE}{Temple University,  Philadelphia, PA 19122 }
\newcommand*{\TEMPLEindex}{36}
\affiliation{\TEMPLE}
\newcommand*{\JLAB}{Thomas Jefferson National Accelerator Facility, Newport News, Virginia 23606}
\newcommand*{\JLABindex}{37}
\affiliation{\JLAB}
\newcommand*{\UTFSM}{Universidad T\'{e}cnica Federico Santa Mar\'{i}a, Casilla 110-V Valpara\'{i}so, Chile}
\newcommand*{\UTFSMindex}{38}
\affiliation{\UTFSM}
\newcommand*{\EDINBURGH}{Edinburgh University, Edinburgh EH9 3JZ, United Kingdom}
\newcommand*{\EDINBURGHindex}{39}
\affiliation{\EDINBURGH}
\newcommand*{\GLASGOW}{University of Glasgow, Glasgow G12 8QQ, United Kingdom}
\newcommand*{\GLASGOWindex}{40}
\affiliation{\GLASGOW}
\newcommand*{\VT}{Virginia Tech, Blacksburg, Virginia   24061-0435}
\newcommand*{\VTindex}{41}
\affiliation{\VT}
\newcommand*{\VIRGINIA}{University of Virginia, Charlottesville, Virginia 22901}
\newcommand*{\VIRGINIAindex}{42}
\affiliation{\VIRGINIA}
\newcommand*{\VIRGINIATECH}{Virginia Polytechnic Institute and State University, Blacksburg, Virginia, 24061}
\newcommand*{\VIRGINIATECHindex}{43}
\affiliation{\VIRGINIATECH}
\newcommand*{\WM}{College of William and Mary, Williamsburg, Virginia 23187-8795}
\newcommand*{\WMindex}{44}
\affiliation{\WM}
\newcommand*{\YEREVAN}{Yerevan Physics Institute, 375036 Yerevan, Armenia}
\newcommand*{\YEREVANindex}{45}
\affiliation{\YEREVAN}
 
\newcommand*{\NOWJLAB}{Thomas Jefferson National Accelerator Facility, Newport News, Virginia 23606}
\newcommand*{\NOWINFNGE}{INFN, Sezione di Genova, 16146 Genova, Italy}
\newcommand*{\NOWUK}{University of Kentucky, Lexington, Kentucky 40506}
\newcommand*{\NOWISU}{Idaho State University, Pocatello, Idaho 83209}
\newcommand*{\NOWODU}{Old Dominion University, Norfolk, Virginia 23529}
\newcommand*{\NOWGABES}{Faculty of Sciences of Gabes, Department of Physics, 6072-Gabes, Tunisia}

\author {M.~Hattawy}
\affiliation{\ANL}
\affiliation{\ORSAY}
\author {N.A.~Baltzell} 
\affiliation{\ANL}
\affiliation{\JLAB}
\author {R.~Dupr\'{e}} 
\email[corresponding author: ]{dupre@ipno.in2p3.fr}
\affiliation{\ANL}
\affiliation{\ORSAY}
\author {K.~Hafidi} 
\affiliation{\ANL}
\author{S.~Stepanyan}
\affiliation{\JLAB}
\author {S.~B\"{u}ltmann} 
\affiliation{\ODU}
\author{R.~De~Vita} 
\affiliation{\INFNGE}
\author {A.~El~Alaoui} 
\affiliation{\ANL}
\affiliation{\UTFSM}
\author {L.~El~Fassi} 
\affiliation{\MISS}
\author{H.~Egiyan}
\affiliation{\JLAB}
% ...

\collaboration{The CLAS Collaboration}
\noaffiliation

\date{\today}
\begin{abstract}
We report on the first measurement of the beam-spin asymmetry in the exclusive 
process of coherent deeply virtual Compton scattering off a nucleus. The 
experiment used the 6 GeV electron beam from the CEBAF accelerator at Jefferson 
Lab incident on a pressurized $^4$He gaseous target placed in front of the CEBAF Large Acceptance 
Spectrometer (CLAS). The scattered electron was detected by CLAS and the photon 
by a dedicated electromagnetic calorimeter at forward angles. To ensure the 
exclusivity of the process, a specially designed radial time projection chamber 
was used to detect the recoiling $^4$He nuclei. We measured beam-spin 
asymmetries larger than those observed on the free proton in the same kinematic 
domain. From these, we were able to extract, in a model-independent way, the 
real and imaginary parts of the only $^4$He Compton form factor, $\cal H_A$. 
This first measurement of coherent deeply virtual Compton scattering on 
the $^4$He nucleus, with a fully exclusive final state via nuclear recoil 
tagging, leads the way toward 3D imaging of the partonic structure of nuclei.
\end{abstract}
\pacs{Valid PACS appear here}

\maketitle 

The generalized parton distribution (GPD) framework offers the opportunity to 
obtain information about the momentum and spatial degrees of freedom of the 
quarks and gluons inside hadrons \cite{Mueller:1998fv,Ji:1996ek,Ji:1996nm,
Radyushkin:1996nd,Radyushkin:1997ki}. In impact parameter space the GPDs are 
indeed interpreted as a tomography of the transverse plane for partons 
carrying a given fraction of the proton longitudinal momentum 
\cite{Burkardt:2000za,Diehl:2002he,Belitsky:2002ep,Burkardt:2005hp}. The most 
promising way to access GPDs experimentally is through the measurement of 
deeply virtual Compton scattering (DVCS), \textit{i.e.}, the hard exclusive 
electroproduction of a real photon on a hadron. While other processes are also known 
to be sensitive to GPDs, the measurement of DVCS is considered the cleanest 
probe and has been the focus of efforts at Jefferson Lab, HERA, and CERN 
\cite{Stepanyan:2001sm,Airapetian,Airapetian:2010nu,Chekanov:2003ya,Aktas:2005ty,Chen:2006na,
MunozCamacho:2006hx,Girod:2007aa,Mazouz:2007aa,Gavalian:2009,Seder:2015,Defurne:2015kxq,
Pisano:2015,Jo:2015ema,Joerg:2016hhs}. The vast majority of these measurements 
focused on the study of the proton and allowed for an extraction of its 
three-dimensional image (for reviews of the field, see~\cite{Goeke:2001tz,
Diehl:2003ny,Ji:2004gf,Belitsky:2005qn,Boffi:2007yc,Guidal:2013rya}). These 
recent developments could also be applicable to nuclei, giving access to novel 
information about nuclear structure in terms of quarks and gluons~\cite{Berger:2001zb,
Cano:2003ju,Guzey:2005ba,Dupre:2015jha}. Such a study of the 3D 
structure of nuclei appears to be especially interesting in light of the large, 
yet unresolved, nuclear effects observed in nuclear parton distribution 
functions~\cite{Geesaman:1995yd,Norton:2003cb,Hen:2016kwk}. The results 
presented in this letter demonstrate the feasability of such an approach and 
constitute the first step toward a tomography of nuclei.

\begin{figure}[tb]
\includegraphics[width=6.5cm]{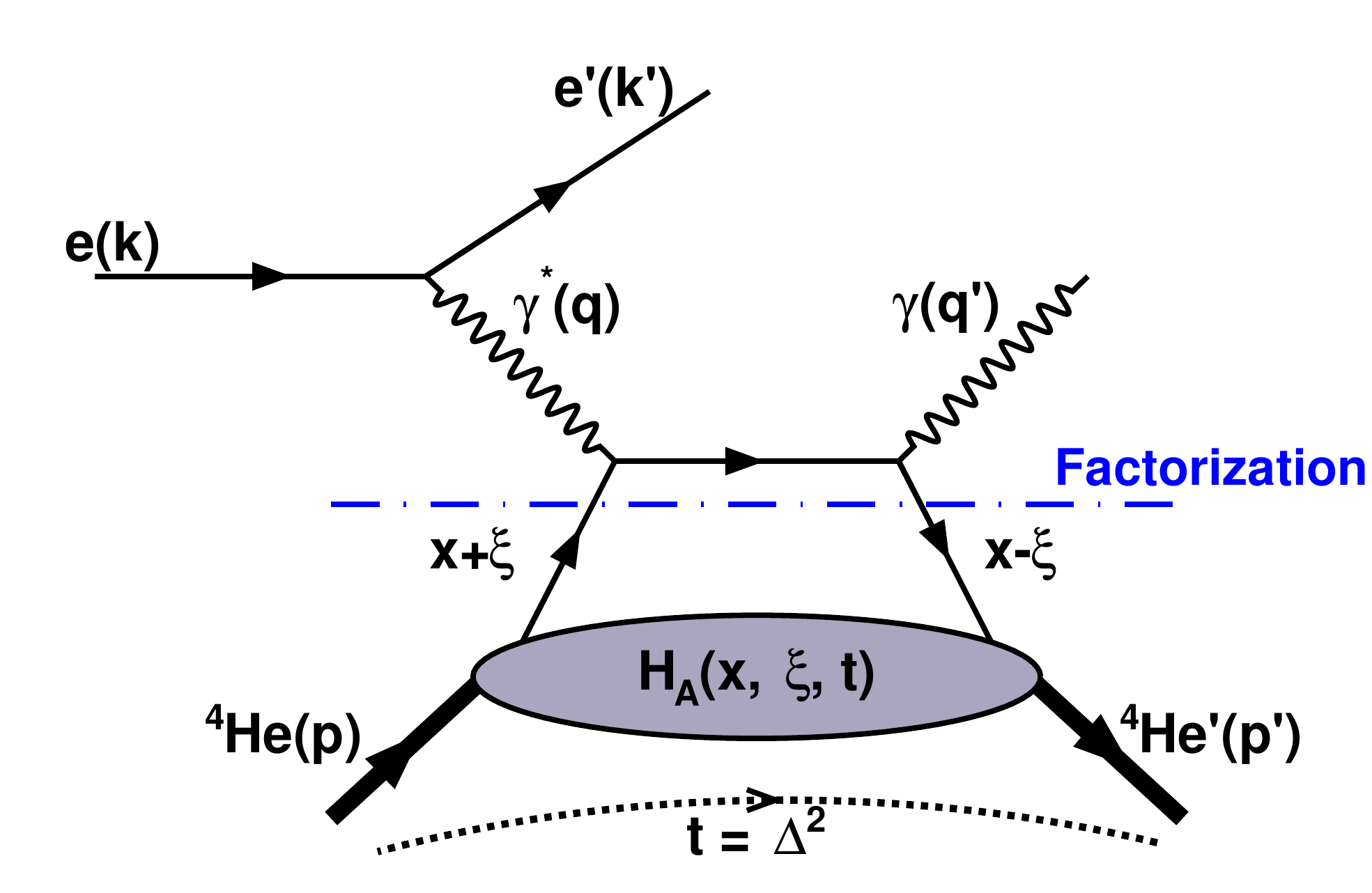}
\caption{Representation of the leading-order handbag diagram of the DVCS 
process off $^4$He.}
\label{fig:diags}
\end{figure}

Figure~\ref{fig:diags} illustrates the handbag diagram for coherent DVCS on 
$^4$He, where the four-vectors of the electrons, photons, and $^4$He are denoted 
by $k/k^\prime$, $q/q^\prime$, and $p/p^\prime$ respectively. For large 
virtual photon 4-momentum squared, $Q^2=-(k-k')^{2}$, and small squared 
momentum transfer, $t=(p-p')^{2}$, the DVCS handbag 
diagram can be factorized into two parts \cite{Freund_Collins,Ji_Osborne}. The 
hard part includes the photon-quark interaction and is calculable in 
perturbative QED. The non-perturbative part is parametrized in terms of GPDs, 
which embed the partonic structure of the hadron. The GPDs depend on the three 
variables $x$, $\xi$, and $t$. $\xi$ relates to the Bjorken variable 
$x_{A}$: $\xi\approx {{x_A}\over{2-x_A}}$, where $x_A=\frac{Q^2}{2M_A\nu}$, 
$\nu$ is the energy of the virtual photon, and 
$M_A$ is the nuclei mass. $x$ is the quark's internal loop momentum 
fraction and cannot be accessed experimentally in DVCS. We in fact measure 
Compton form factors (CFF), which are complex quantities defined as:
\begin{align}
\begin{split}
\Re e(&\mathcal{H}_{A}) = \\
    &\mathcal{P} 
\int_{0}^{1}dx[H_A(x,\xi,t)-H_A(-x,\xi,t)] \, C^{+}(x,\xi), 
\end{split} \\
\Im m(&\mathcal{H}_{A}) = - \pi ( H_A(\xi,\xi,t)-H_A(-\xi,\xi,t)),
\end{align}
with $H_A$ a GPD, $\mathcal{P}$ 
the Cauchy principal value integral, and a coefficient function $C^{+}= 
\frac{1}{x-\xi} + \frac{1}{x+\xi}$.

Until now, the only available data on nuclear DVCS were from the HERMES 
experiment \cite{Airapetian:2010nu}. In this experiment, the exclusivity of 
the reaction was obtained through kinematic cuts using only the measured 
scattered electron and real photon. This measurement was performed on a large 
set of nuclei ($^4$He, $^{14}$N, $^{20}$Ne, $^{85}$Kr and $^{131}$Xe), 
but contamination from incoherent processes can be suspected to influence the results 
significantly~\cite{Guzey:2003jh}. The direct detection of the recoil nucleus 
can however guarantee that the nucleus remains intact. 

The $^4$He nucleus is an ideal experimental target for nuclear DVCS, as it is 
light enough to be detected by our experimental setting, while it is subject 
to interesting nuclear effects~\cite{JSeely}. Its spin-zero also leads to an 
important simplification, as a spin-zero hadron is parametrized by only one 
chiral even GPD ($H_{A}(x,\xi,t)$) at leading-twist, while four GPDs arise for 
the spin-$\frac{1}{2}$ nucleon. This significantly simplifies the interpretation of the 
data and allows a model independent extraction of the $^4$He CFF 
($\mathcal{H}_{A}$) presented at the end of this letter. 

The CEBAF Large Acceptance Spectrometer (CLAS) in Hall B at Jefferson 
Lab~\cite{Mecking:2003zu} has been previously supplemented with an inner 
calorimeter (IC) and a solenoid magnet to measure DVCS observables on the nucleon
\cite{Girod:2007aa,Gavalian:2009,Seder:2015,Pisano:2015,Jo:2015ema}. The IC 
extended the photon detection acceptance of CLAS to polar angles as low as 
4$^{\circ}$. The 5-T solenoid magnet acted as a guiding field for the 
low-energy M\o{}ller electrons that were absorbed in a  
heavy shield placed around the beam line. 

In the kinematic range of the present experiment, the recoil $^4$He nuclei 
have low momentum, averaging 300 MeV. CLAS could not detect such low-energy $\alpha$ 
particles, so in order to ensure the exclusivity of the measurement, we built a 
small and light radial time projection chamber (RTPC). The 
RTPC was a $20$-cm-long cylinder with a diameter of $15$ cm, positioned in the 
solenoid magnet. In the center of the RTPC was the target cell, a $25$-cm-long 
and $6$-mm-diameter Kapton tube with $27$-$\mu$m-thick walls filled with 
gaseous $^4$He at 6~atm (see~\cite{Dupre:2017upj} for a detailed description of 
the RTPC and its performances). The experiment (E08-024)~\cite{Hafidi:2008pr} collected
data over 40 days at the end of 2009 using a 
nearly 100\% duty factor, longitudinally polarized electron beam (83.7$\pm 3.5 \%$ 
polarization~\cite{Perrin:thesis}) at an energy of 6.064 GeV. The RTPC was calibrated specifically 
for the detection of $^4$He nuclei using elastic scattering 
($e^4$He$\to e^\prime$$^4$He) with a 1.2~GeV electron beam.
 
To identify coherent DVCS events, we first selected events where one electron, 
one $^4$He, and at least one photon were detected in the final state. Electrons 
were identified using their measured momentum, light yield, time, and energy obtained from 
the CLAS drift chambers, \v{C}erenkov counters, scintillator counters, 
and electromagnetic calorimeters, respectively. The recoiling $^4$He nuclei 
were identified in the RTPC using time and energy-loss cuts for tracks in the 
fiducial region~\cite{Hattawy:thesis}. In addition, we applied a vertex-matching 
cut to ensure that the 
electron and helium nucleus originated from a common reaction vertex in the 
target cell. The photons were 
detected in either the IC or the CLAS electromagnetic calorimeters. Note that 
even though the DVCS reaction has only one real photon in the final state, 
events with more than one good photon were not discarded at this stage. These
were mainly caused by accidental coincidences of soft photons and did not directly
affect this measurement, as only the most energetic photon of an 
event was considered a DVCS photon candidate.
This prescription however slightly increased the corrections associated with
the $\pi^0$ and the accidental backgrounds discussed below. 

\begin{figure}[tb]
\centering
\hspace{-0.45cm}
\includegraphics[width=9cm]{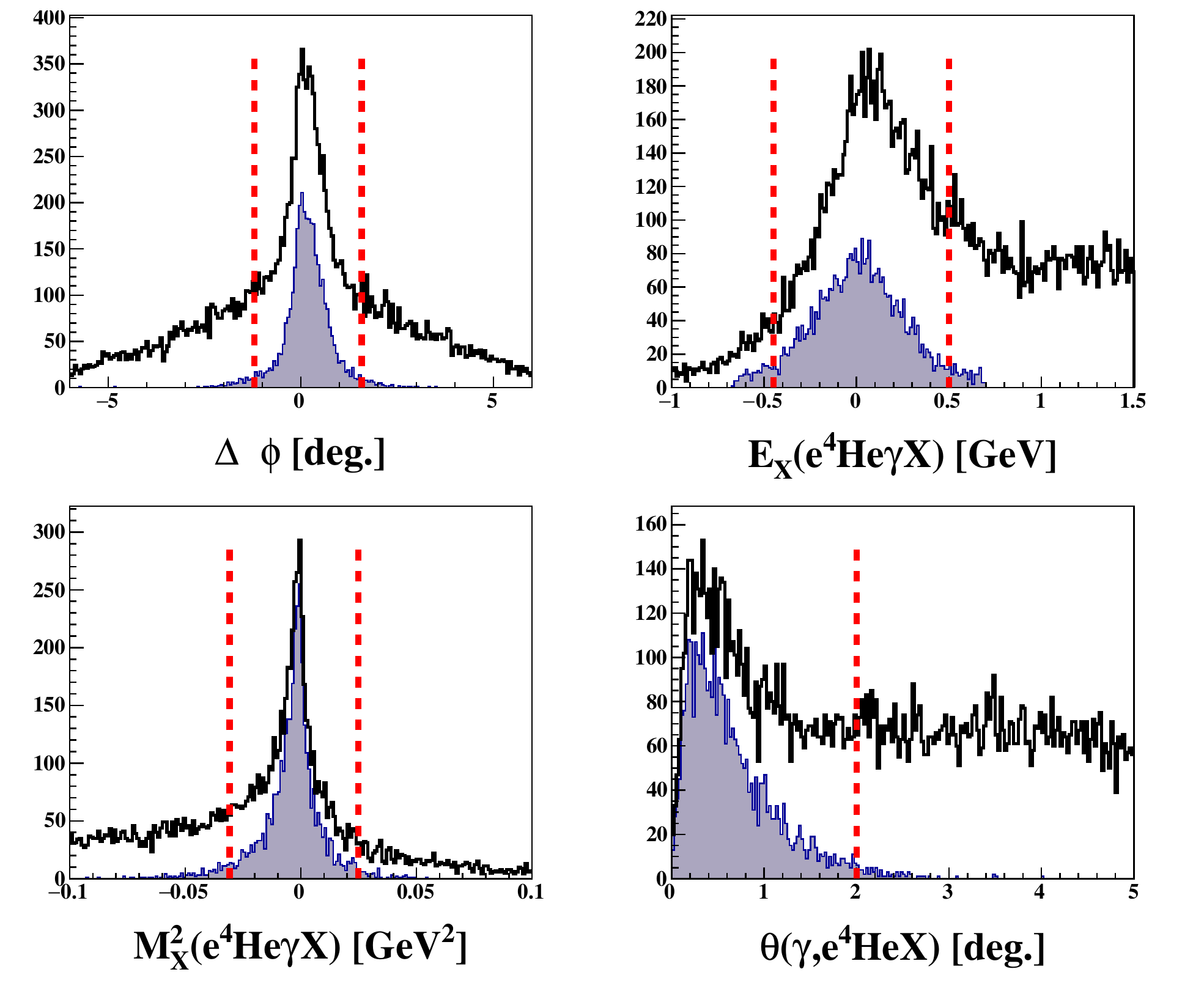}
\caption{Four of the six coherent DVCS exclusivity cuts. The black 
distributions represent the initial candidate events, while the shaded 
distributions represent those that passed all of the exclusivity cuts except 
the quantity plotted. The vertical red lines represent the applied cuts.
The distributions from left to right and from top to bottom are: coplanarity 
angle $\Delta \phi$, missing energy $E_X$, missing-mass-squared $M_X^2$, and 
the cone angle $\theta$ between the measured photon and the missing momentum 
of the $e^\prime{^4}$He$^\prime$ system.}
\label{fig:kin-cuts}
\end{figure}

\begin{figure}[tb]
\hspace{-0.45cm}
\includegraphics[width=9cm]{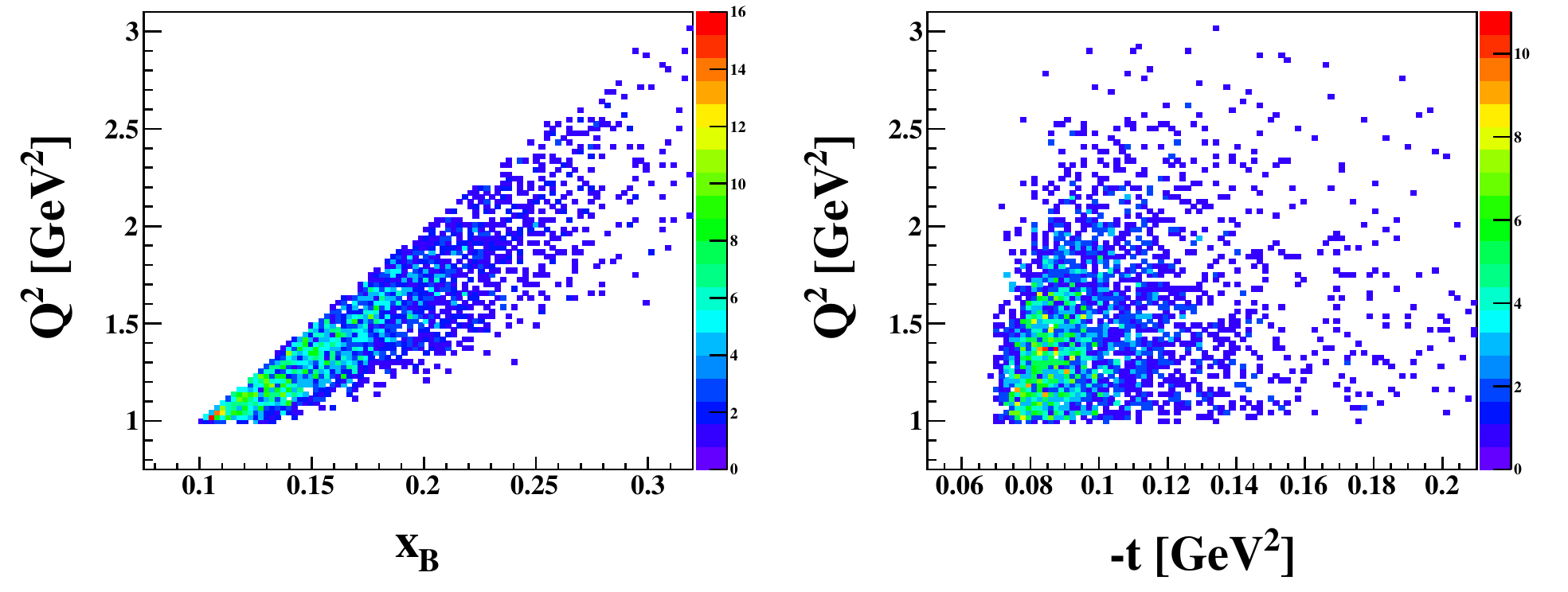}
\caption{Coherent DVCS event distributions for $Q^2$ after exclusivity cuts. 
The distributions are shown as a function of Bjorken variable $x_B$ (left) 
and as a function of squared-momentum transfer $-t$ (right).}
\label{fig:kin-coverage}
\end{figure}

We selected events with $Q^{2}$ greater than 1~GeV$^{2}$ for which the DVCS handbag 
diagram is believed to be dominant. Then the exclusivity of the reaction was ensured by 
applying a set of cuts on the following kinematic variables: the coplanarity 
angle $\Delta\phi$ between the ($\gamma,\gamma^*$) 
and ($\gamma^*$,$^4$He$'$) planes, the missing energy, mass, and transverse
momentum of the $e'^4$He$'\gamma$ system, the missing mass squared of 
the $e'^4$He$'$ system, and the angle $\theta$
between the measured photon and the missing momentum of the $e'^4$He$'$ system.  
The experimental data for the most relevant exclusivity variables and applied
cuts are shown in Fig.~\ref{fig:kin-cuts} (see \cite{Hattawy:thesis} for 
additional details). We also 
rejected events where a $\pi^0$ was identified by the invariant mass of 
two photons. About 3200 events passed all these requirements, their kinematic 
distributions\footnote{We use here and for other results 
$x_B =\frac{Q^2}{2M_N\nu}$ with $M_N$ the proton mass instead of $x_A$.
This makes it easier to compare these results with the proton DVCS data
available in the literature.} are shown in Fig.~\ref{fig:kin-coverage}.

We identified two main backgrounds, accidental coincidences and exclusive coherent
$\pi^0$ production. The accidentals have particles originating from different events,
and we estimated their contribution to be 4.1\% of the data sample. We evaluated this 
contribution by selecting events passing all the cuts but with the scattered electron and 
$^4$He originating from different vertices. The $\pi^0$ production can
be mistaken for DVCS when one of the two photons from the $\pi^0$ 
decay is produced at low-energy in the laboratory frame and remains undetected.  
To estimate the effect of this contamination, we developed an event generator 
tuned on the experimental yield of exclusive $\pi^0$ with two photons measured. 
We used this generator together with a GEANT3 simulation of our 
detectors to estimate the ratio of the number of $\pi^0$ events where the 
two photons are detected to those that are misidentified as DVCS events. This 
ratio is then multiplied by the measured yield of exclusive $\pi^0$ events to 
correct the DVCS data. Depending on the kinematics, we found contaminations of 
2 to 4\%. 

In this work, the physics observable extracted using coherent DVCS events is
the beam-spin asymmetry, $A_{LU}$. On an unpolarized target, $A_{LU}$ is 
defined as the difference of cross sections for the reaction with opposite beam 
helicities normalized to the total cross section:
  \begin{equation}
  A_{LU} = \frac{d^{4}\sigma^{+} - d^{4}\sigma^{-} }
                {d^{4}\sigma^{+} + d^{4}\sigma^{-}},
    \label{BSA_equation}
  \end{equation}
where $d^{4}\sigma^{\pm}$ is the DVCS differential cross 
section for positive (negative) beam helicity. 

In this ratio, luminosity normalization and
detector efficiencies largely cancel and $A_{LU}$ can be 
extracted from the reaction yields for the two helicities ($N^{\pm}$):
\begin{equation}
A_{LU} = \frac{1}{P_{B}} \frac{N^{+} - N^{-}}{N^{+} + N^{-} },
\end{equation}
where $P_{B}$ is the degree of longitudinal polarization of the incident electron beam.

There is an additional process contributing to the same final state as the 
DVCS, the so-called Bethe-Heitler (BH) process, where the real photon is 
emitted by the incoming or the outgoing lepton. The DVCS and BH processes are 
indistinguishable experimentally and the amplitude of electroproduction of real
photons includes a sum of the amplitudes of these two processes. The BH 
amplitude depends on the target elastic form factors, which are well known in this
kinematic region, while the DVCS amplitude depends on the GPDs we are trying to
measure. For our kinematics, the cross section of real photon 
electroproduction is dominated by the BH contribution, while the DVCS 
contribution is very small. However, the DVCS contribution is
enhanced in the observables sensitive to the interference term, {\it e.g.} 
$A_{LU}$. The three terms entering the cross section calculation,
the squares of the BH and DVCS amplitudes and their interference term, depend on the
azimuthal angle $\phi$ between the $(e,e^\prime)$ and $(\gamma^*,^4$He$^\prime)$ planes,
as shown for the nucleon in Ref.~\cite{Belitsky:2001ns} and for the spin-zero targets
in Refs.~\cite{Kirchner:2003wt,Belitsky:2008bz}. Based on this work, $A_{LU}$ 
for a spin-zero hadron can be expressed at leading-twist as
\begin{equation}
\begin{split}
&A_{LU}(\phi) = \\
&\frac{\alpha_{0}(\phi) \, \Im m(\mathcal{H}_{A})}
{\alpha_{1}(\phi) + \alpha_{2}(\phi) \, \Re e(\mathcal{H}_{A}) + \alpha_{3}(\phi) \, 
\big( \Re e(\mathcal{H}_{A})^{2} + \Im m(\mathcal{H}_{A})^{2} \big)}.
\end{split}
\label{eq:A_LU-coh}
\end{equation}
Explicit expressions of the kinematic factors $\alpha_i$ are derived 
from expressions in
Ref.~\cite{Belitsky:2008bz} and are functions of Fourier harmonics in the 
azimuthal angle $\phi$, the nuclear form factor $F_A(t)$ and kinematical
factors. Using the different $\sin(\phi)$ and $\cos(\phi)$ contributions, 
both the imaginary and real parts of $\mathcal{H}_{A}$ can be extracted unambiguously
by fitting the $A_{LU}(\phi)$ distribution.

\begin{figure}[tb]
   \centering
\includegraphics[width=8.5cm]{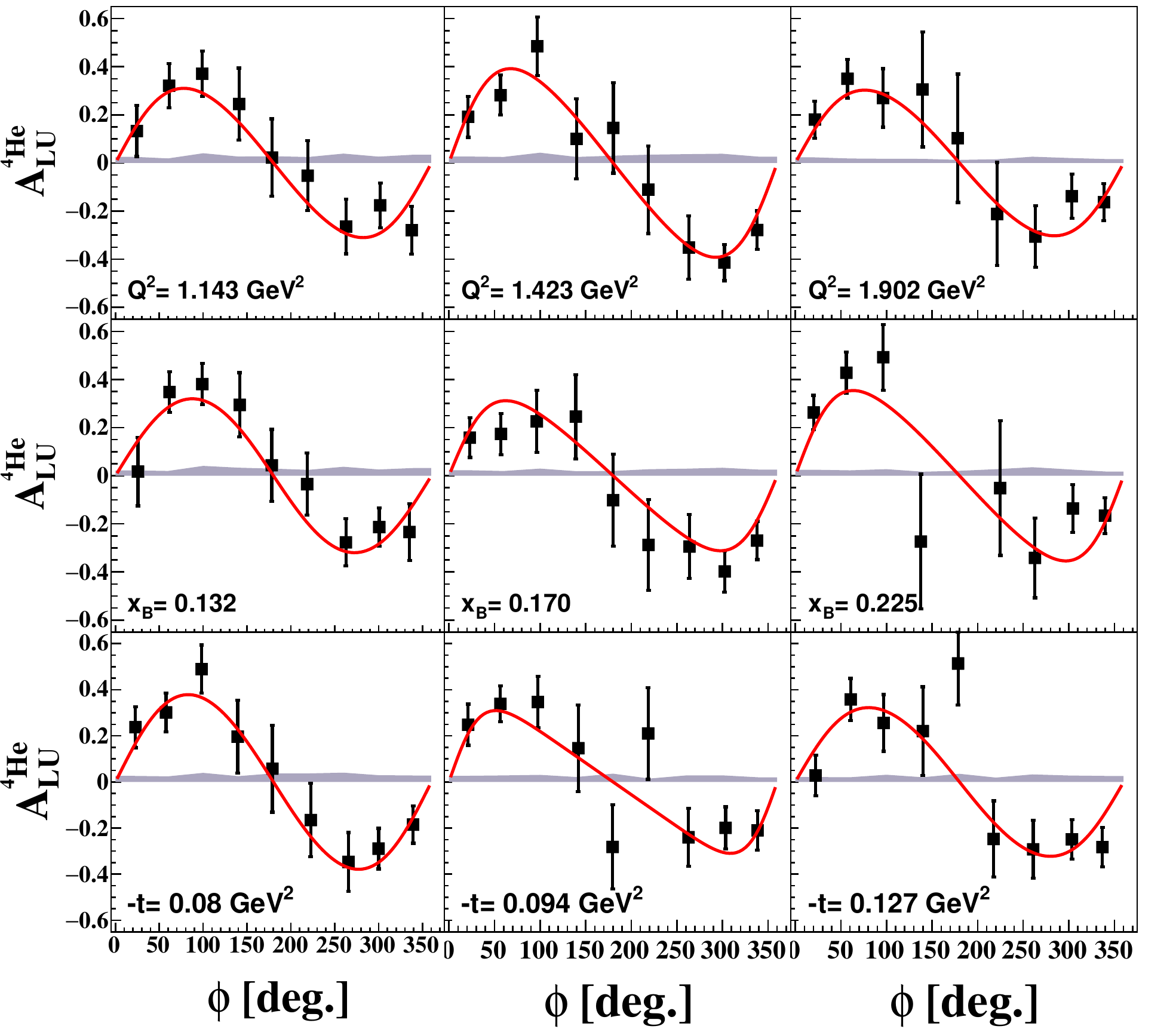}
\caption{$A_{LU}$ as a function of azimuthal angle $\phi$. Results are presented
   for different $Q^{2}$ bins (top panel), $x_{B}$ bins (middle panel), and $t$ 
   bins (bottom panel).  The error bars represent the statistical 
   uncertainties. The grey bands represent the systematic uncertainties, 
   including the normalization uncertainties. The red curves are the results of 
   fits with Eq.~(\ref{eq:A_LU-coh}).}
\label{fig:alu}
\end{figure}

We present in Fig.~\ref{fig:alu} $A_{LU}$ as a function of azimuthal angle 
$\phi$ and the kinematical variables $Q^2$, $x_B$, and $t$. Due to limited 
statistics, these latter variables are studied separately with a two-dimensional 
data binning. The curves on 
the plots are fits using the function presented in Eq.~(\ref{eq:A_LU-coh}), 
where the real and imaginary parts of the CFF $\mathcal{H}_{A}$ are the only 
free parameters. 

Studies of systematic uncertainties showed that the main contributions 
come from the choice of DVCS exclusivity cuts (8\% systematic uncertainty) and the 
large binning size (5.1\%). These values are relative and quoted for $A_{LU}$
at $\phi=90^\circ$. Added quadratically, the total systematic uncertainty
is about 10\% at $90^\circ$ (or 0.03, absolute), which is significantly smaller
than the statistical uncertainties at all kinematical bins. 

\begin{figure*}[tb]
\includegraphics[width=16cm]{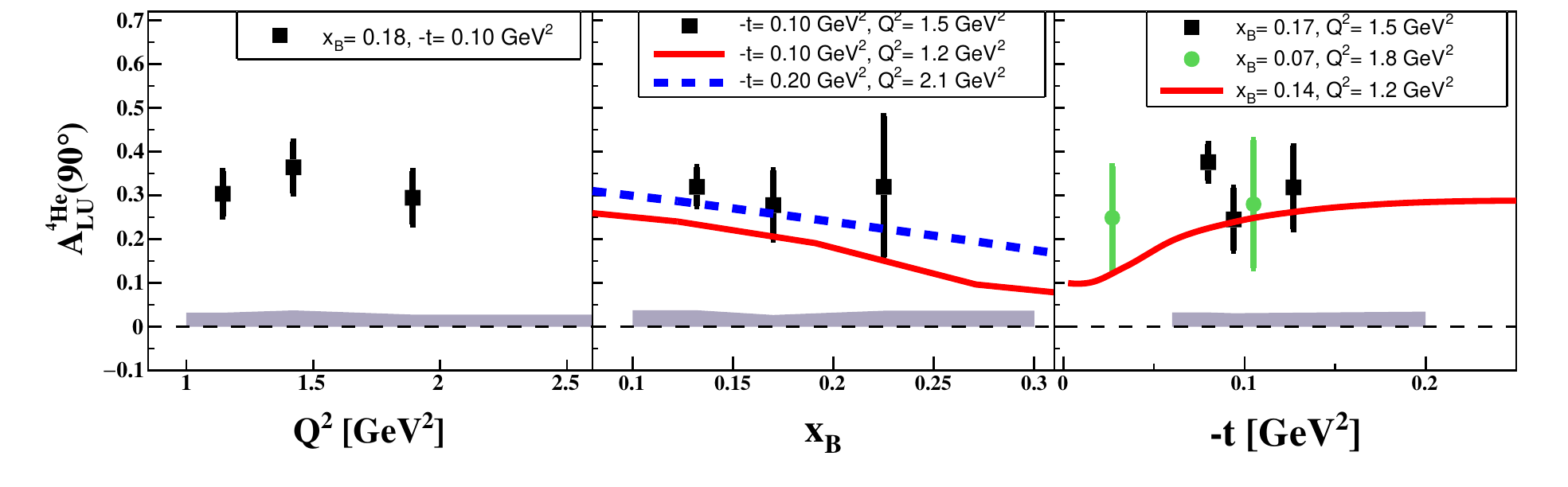}
\caption{The $Q^{2}$ (left), $x_{B}$ (middle), and $t$-dependencies (right) of
   $A_{LU}$ at $\phi$~=~90$^{\circ}$ (black squares). On the 
   middle plot: the full-red and the dashed-blue curves are theoretical 
   calculations from \cite{simonetta_2}. On the right: the green circles are 
   the HERMES $-A_{LU}$ (a positron beam was used) inclusive measurements 
   \cite{Airapetian}, and the curves are the theoretical calculations 
   from \cite{simonetta_2}. The grey bands represent the systematic errors.}
\label{fig:alu90}
\end{figure*}

In Fig.~\ref{fig:alu90}, the $Q^2$, $x_{B}$, and $t$-dependencies of the fitted 
$A_{LU}$ at $\phi$~=~90$^{\circ}$ are shown. The comparison to HERMES data
shows that we obtain the same sign, but the size of the error bars and the difference
of kinematics do not permit to say much more. The $x_{B}$ and $t$-dependencies 
are also compared to theoretical calculations by S.~Liuti and K.~Taneja 
\cite{simonetta_2}. The model accounts for the effect of the nucleon virtuality 
(off-shellness) on the quark distribution. The calculations are at slightly 
different kinematics than the data but still allow us to draw some conclusions. The 
model appear to predict smaller asymmetries than observed. 
The difference may arise from the theoretical uncertainty in the determination 
of the crossing point where the parton nuclear distribution becomes larger than the 
nucleon one, and reverses the sign of the nuclear effect.

\begin{figure}[tb]
\includegraphics[width=8.5cm]{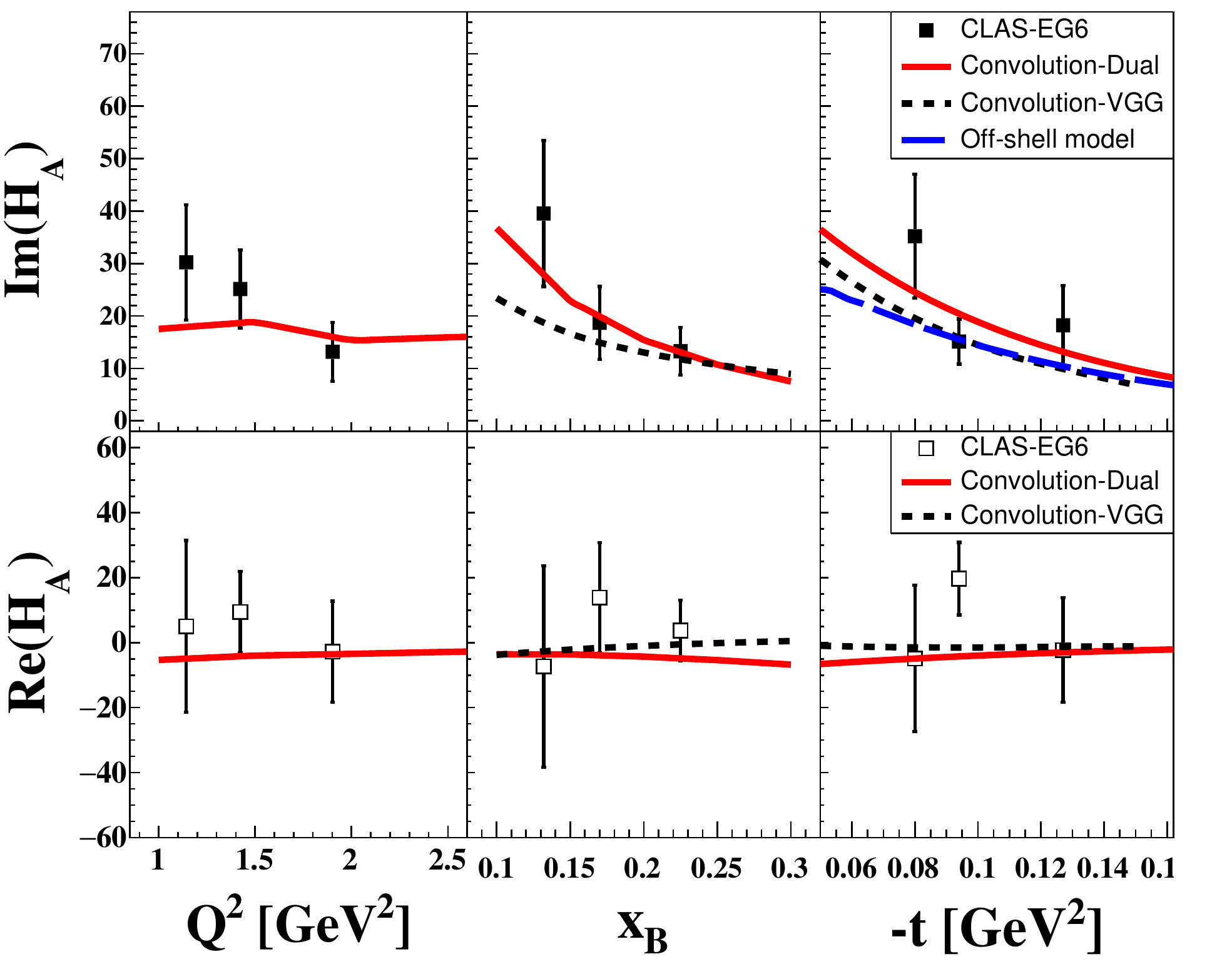}
\caption{The leading-twist model-independent extraction of the imaginary (top panel) and
real (bottom panel) parts of the $^4$He CFF $\mathcal{H}_A$, as a function of
$Q^{2}$ (left panel), $x_B$ (middle panel), and $t$ (right panel). The full red 
curves are the calculations based on the convolution model from Ref.~\cite{Vadim_priv}.
The black-dashed curves are calculations from the same 
model using different GPDs for the nucleons (VGG model~\cite{Guidal_priv}). The 
blue long-dashed curve on the top-right plot is from
the off-shell model as described in Ref.~\cite{GonzalezHernandez:2012jv}.}
\label{fig:CFF_HA}
\end{figure}

The $Q^2$, $x_B$, and $t$ dependencies of the $^4$He CFF $\mathcal{H}_A$ 
extracted from the fit to the azimuthal dependence of $A_{LU}$ are shown in 
Fig.~\ref{fig:CFF_HA}. The curves on the graphs are model calculations, labelled 
{\it convolution} and {\it off-shell}. In the convolution model 
\cite{Vadim_priv}, the nucleus is assumed to be composed of non-relativistic 
nucleons, each interacting independently with the probe. The Convolution-Dual 
model is based on nucleon GPDs from the dual parametrization 
\cite{Guzey:2006xi}, where the Convolution-VGG uses nucleon GPDs  from the VGG 
model and is based on the double distributions ansatz \cite{DD_model}. The 
off-shell model is the same as in Fig.~\ref{fig:alu90} using a more recent
GPD model for the nucleon \cite{GonzalezHernandez:2012jv}.

The results in Fig.~\ref{fig:CFF_HA} show that the extraction of the CFF
from the $A_{LU}$ is possible without model-dependent assumptions beyond
leading-twist and leading-order dominance. The amplitude and the dependencies 
observed as a function of 
$Q^{2}$, $x_B$, and $t$ are in agreement with the theoretical expectations. One 
can see a difference between the precision of the extracted imaginary and real 
parts, which is is due to $\alpha_2$ being much smaller than $\alpha_1$ in
Eq.~(\ref{eq:A_LU-coh}). While the precision of this measurement is not at a
sufficient level to discriminate between the models, these results demonstrate
the possibility of extracting the CFF of a spin-0 target directly from a $A_{LU}$
measurement.

In summary, we have presented the first measurement of the beam-spin asymmetry
of exclusive coherent DVCS off $^4$He using the CLAS spectrometer supplemented 
with a RTPC. This setup allowed detection of the 
low-energy $^4$He recoils in order to ensure an exclusive measurement of the
coherent DVCS process. The azimuthal dependence of the measured $A_{LU}$ has 
been used to extract, in a model-independent way, the real and the imaginary 
parts of the $^4$He CFF, $\mathcal{H}_A$. The extracted CFF is in  
agreement with predictions of the available models. This first fully exclusive 
experiment opens new perspectives for studying nuclear structure with the GPD 
framework and paves the way for future measurements at JLab using 12 GeV CEBAF 
and upgraded equipment.

The authors thank the staff of the Accelerator and Physics Divisions at the 
Thomas Jefferson National Accelerator Facility who made this experiment 
possible. This work was supported in part by the Chilean Comisi\'on Nacional 
de Investigaci\'on Cient\'ifica y Tecnol\'ogica (CONICYT), the Italian 
Instituto Nazionale di Fisica Nucleare, the French Centre National de la 
Recherche Scientifique, the French Commissariat \`a l'Energie Atomique, the 
U.S. Department of Energy, the UK Science and Technology Facilities Council 
(STFC), the Scottish Universities Physics Alliance (SUPA), the National 
Research Foundation of Korea, and the Office of Research and Economic 
Development at Mississippi State University.  The author (M.~Hattawy) also 
acknowledges the support of the Consulat G\'en\'eral de France \`a J\'erusalem. The Southeastern Universities Research Association operates the Thomas 
Jefferson National Accelerator Facility for the United States Department of 
Energy under contract DE-AC05-06OR23177.

\end{document}